\documentclass[aps,prb,showpacs,superscriptaddress,twocolumn]{revtex4}
\usepackage{graphicx}
\usepackage{amsmath}
\usepackage{bm}
\usepackage{xcolor}

\begin{document}

\title{Geometric Effect on Quantum Anomalous Hall States in Magnetic Topological Insulators}

\author{Yanxia Xing}
\affiliation{Beijing Key Laboratory of Nanophotonics and Ultrafine Optoelectronic Systems, School of Physics, Beijing Institute of Technology, Beijing 100081, China}

\author{Fuming Xu}
\affiliation{Shenzhen Key Laboratory of Advanced Thin Films and Applications, College of Physics and Energy, Shenzhen University, Shenzhen 518060, China}

\author{Qing-feng Sun}
\affiliation{International Center for Quantum Materials, School of Physics, Peking University, Beijing 100871, China}
\affiliation{Collaborative Innovation Center of Quantum Matter, Beijing 100871, China}

\author{Jian Wang}
\email{jianwang@hku.hk}
\affiliation{Department of Physics and the Center of Theoretical and Computational Physics, The University of Hong Kong, Pokfulam Road, Hong Kong, China}

\author{Yu-gui Yao}
\email{ygyao@bit.edu.cn}
\affiliation{Beijing Key Laboratory of Nanophotonics and Ultrafine Optoelectronic Systems, School of Physics, Beijing Institute of Technology, Beijing 100081, China}

\begin{abstract}
An intriguing observation on the quantum anomalous Hall effect(QAHE) in magnetic topological insulators (MTIs) is the dissipative edge states, where quantized Hall resistance is accompanied by nonzero longitudinal resistance. We numerically investigate this dissipative  behavior of QAHE in MTIs with a three-dimensional tight-binding model and non-equilibrium Green's function formalism. It is found that, in clean samples, the geometric mismatch between the detecting electrodes and the MTI sample leads to additional scattering in the central Hall bar, which is similar to the effect of splitting gates in the traditional Hall effect. As a result, while the Hall resistance remains quantized, the longitudinal resistance deviates from zero due to such additional scattering. It is also shown that external magnetic fields as well as disorder scattering can suppress the dissipation of the longitudinal resistance. These results are in good agreement with previous experimental observations and provide insight on the fabrication of QAHE devices.
\end{abstract}

\pacs{
73.23.Ad	
73.40.Cg	
73.50.Bk	
73.25.+i}	

\maketitle

\section{introduction}
A topological insulator is a new quantum state of matter,\cite{Rev.Mod.Phys.82.3045-3067.Hasan.2010,PhysicalReviewB78.195424.Qi.2008,PhysicalReviewLetters98.106803.Fu.2007} in which a couple of helical or chiral conducting boundary states reside in the bulk insulating gap stably. Novel quantum spin Hall effect (QSHE)\cite{Science301.1348-1351Murakami2003,Science314.1757-1761Bernevig2006,Science318.766-770Konig2007,PhysicalReviewLetters100.Liu2008} and quantum anomalous Hall effect (QAHE)\cite{Phys.Rev.Lett.61.2015.Haldane.1988,PhysicalReviewB82.Qiao2010,PhysicalReviewB83.Tse2011,PhysicalReviewBRen2017,PhysicalReviewLettersQiao2016} have been suggested in topological insulator systems. Comparing to the QSHE, QAHE is more promising in future device applications since it is robust against magnetic disorder scattering. QAHE was originally discussed in various two-dimensional models with broken time-reversal symmetry, especially the honeycomb lattice\cite{Phys.Rev.Lett.61.2015.Haldane.1988} and monolayer/bilayer graphene systems.\cite{PhysicalReviewB82.Qiao2010,PhysicalReviewB83.Tse2011} Recently, QAHE was theoretically proposed in magnetic topological insulators (MTIs).\cite{Phys.Rev.Lett.101.146802.Liu.2008,Science329.61-64.Yu.2010} Soon after, the existence of QAHE in three-dimensional(3D) magnetically doped $(Bi,Sb)_2Te_3$ films was confirmed by a series of experiments.\cite{Science340.167-170.Chang.2013,PhysicalReviewLetters113.Kou.2014,PhysicalReviewLetters114.187201.Bestwick.2015,NatureMaterials14.473-477.Chang.2015}

\begin{figure}
\includegraphics[width=8.5cm,totalheight=7cm]{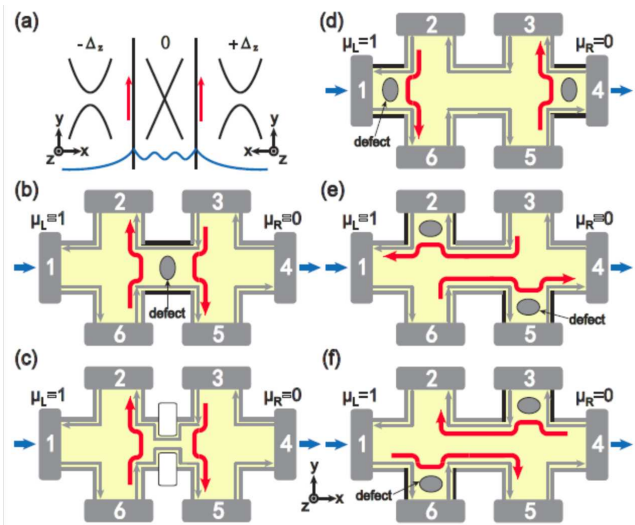}
\caption{ (Color online)
Panel (a): The gaped Dirac-like top and bottom surface with opposite-signed effective mass unidirectional chiral edge states and gapless Dirac-like side surface in the presence of exchange fields. Panel (b): The additional scattering appears in the Hall bar due to the non-ideal contact between the terminal leads and central scattering region, as the effect of splitting gates in panel (c). Panels (d-f): The additional scattering in the entrances of leads induces the floating chiral edge states. The additional scattering induced by the non-ideal contact is similar to the effect of defects (the gray solid ellipses).}
\end{figure}

In a 3D TI, the conducting surface holds massless Dirac Fermions. In the presence of ferromagnetic exchange field along $z$-direction, the time-reversal symmetry is broken and the nontrivial gap is opened in the top and bottom gapless surfaces with opposite-signed effective mass shown in Fig.1(a), which induces non-dissipative chiral edge states and consequently the ideal QAHE with perfectly quantized Hall resistance $h/e^2$ and zero longitudinal resistance. However, in the experimental observations,\cite{Science340.167-170.Chang.2013,PhysicalReviewLetters113.Kou.2014,PhysicalReviewLetters114.187201.Bestwick.2015}
the chiral edge states of QAHE are usually dissipative, where the Hall resistance $\rho_{xy}$ is quantized but the longitudinal resistance $\rho_{xx}$ significantly deviates from zero. The origin of this derivation attracts intensive research interests and various mechanisms has been suggested, for instance, thermally activated carriers,\cite{PhysicalReviewLetters114.187201.Bestwick.2015} the alignment of exchange fields,\cite{NatureMaterials14.473-477.Chang.2015} the existence of extra nonchiral edge states,\cite{PhysicalReviewLetters111.086803.Wang.2013} etc. All these delicate interpretations are based on ideal 2D models. However, an MTI thin film is a quasi-2D system in $x$-$y$ plane but has finite thickness in $z$-direction. Different from Refs.[\citenum{Phys.Rev.Lett.61.2015.Haldane.1988,PhysicalReviewB82.Qiao2010}] and Refs.[\citenum{PhysicalReviewB85.045445.Jiang.2012,PhysicalReviewLetters111.136801.Wang.2013}], in which QAHE is considered in ideal 2D systems or superposition of ideal 2D systems, QAHE in MTI thin films is contributed by two half-integer quantum Hall conductances from the top and bottom Dirac-like surfaces with opposite-signed effective mass.\cite{Phys.Rev.B84.085312.Chu.2011,NewJournalofPhysicsXing2018a} The edge states of QAHE actually propagate through side surfaces of MTI films. Therefore, the geometric structure of the side surface along $z$-direction, as a result of nano-fabrication, is important in the formation of QAHE in MTI thin films.

In this work, we study the effect of geometric mismatch on QAHE in MTIs. As shown in Fig.1(b), the imperfection of the detecting electrodes induces geometric barriers in the Hall bar (the thick black lines). Then the chiral edge state of QAHE is backscattered near the entrance of the central Hall bridge (the red arrowed lines), \emph{as though} it is scattered by defects (gray ellipse) in the Hall bridge. This additional scattering is similar to the effect of splitting gates in the Hall effect, in which part of edge states is backscattered due to the pinchoff of the Hall bridge depicted in Fig.1(c).\cite{S.Datta.1995} Using the effective 3D tight-binding Hamiltonian and the non-equilibrium Green's function(NEGF) method, we numerically calculate the longitudinal resistance $\rho_{xx}$ and Hall resistance $\rho_{xy}$ of this six-terminal system. Assuming the probability of the additional scattering in Hall bar is $p$, the longitudinal resistance and Hall resistance are found to be $\rho_{xx}\approx p(h/e^2)$ and $\rho_{xy}=h/e^2$, respectively. This result provides reasonable explanations on the deviation from ideal QAHE in recent experiments.\cite{Science340.167-170.Chang.2013,PhysicalReviewLetters113.Kou.2014,PhysicalReviewLetters114.187201.Bestwick.2015,NatureMaterials14.473-477.Chang.2015} It is also found that additional scattering due to the non-ideal contact can be suppressed and finally eliminated by strong magnetic fields, which is in agreement with experimental observations. Meanwhile, Anderson-type disorders induced by magnetic doping can also suppress the dissipation of $\rho_{xx}$.

The paper is organized as follows. In Sec.II, we present theoretical analysis on the dissipative edge states of QAHE in MTIs. In Sec.III, we numerically calculate $\rho_{xx}$ and $\rho_{xy}$ of a 3D MTI system using NEGF method, through which we present how the theoretical analysis is related to the experimental observation. Finally, a summary of our work is presented in Sec.IV.

\section{Theoretical analysis}

In an MTI film, the helical surface states are gaped in the top and bottom surfaces and gapless in the side surfaces, as shown in Fig.1(a). When the Fermi energy is located in the  surface energy gap induced by the exchange field $M_z$, the surface states are localized. Therefore, only the chiral zero mode\cite{Shen.2012} propagates along the surface boundaries [the red arrows in Fig.1(a)], through the gapless side surfaces [the middle region in Fig.1(a)]. In this case, the chiral states [the gray arrows in Fig.1(b)-(f)] should dominate the transport in the six-terminal Hall system, leading to the $\nu=1$ QAH state. In the experimental setup, the metallic electrodes are pinched to the MTI film, which could induce mismatch between electrodes and the central MTI sample. Considering the geometric structure of the interface between leads and the central scattering region, we examine the following two cases:

(1) The thickness $N$ of terminal leads is larger than that of the central region, i.e., $N_l>N_c$. In this case, electrons are scattered by the boundary of the lead [the thick black lines in Fig.1(d)-(f)]. As a result, additional scattering for floating edge states occurs near the entrances of six terminals as shown in Fig.1(d)-(f). In Fig.1(d) and Fig.1(e), terminal $2$ 'sees' only the channels originating from terminal $3$, so $V_2=V_3$. In Fig.1(f), terminal $3$ 'sees' only the channels coming from terminal $4$ and $V_3=V_4$, and terminal $2$ 'sees' the channels originating from both terminal $3$ and terminal $4$. Since $V_3=V_4$, we have $V_2=V_3=V_4$. It means that there is no voltage drop along the longitudinal direction. In another word, the longitudinal resistance $\rho_{xx}\approx 0$ in the case of $N_l>N_c$.

Besides the intuitive analysis, we can also derive the expressions of $\rho_{xx}$ and $\rho_{xy}$ from the Landauer-B\"{u}ttiker formula.\cite{S.Datta.1995} The current flowing from terminal $m$ to the central scattering region can be calculated as
\begin{equation}
J_m=\frac{e^2}{h}\sum_nT_{mn}(V_m-V_n)\label{Landauer}
\end{equation}
where $m,n=1,2,...,6$. $V_m$ is the voltage on terminal $m$, and $T_{mn}$ is the transmission coefficient from terminal $n$ to terminal $m$. In Eq.(\ref{Landauer}), we have used the gauge invariance condition $\sum_nT_{mn}=\sum_nT_{nm}$. In the measurement of QAHE, a bias $V$ is applied across terminal $1$ and terminal $4$ to inject current, as shown in Fig.1. The other terminals $2$, $3$, $5$ and $6$ are voltage probes and have zero currents, i.e., $J_2=J_3=J_5=J_6=0$. Using the boundary conditions $V_1 = V$, $V_4 = 0$, $J_2=J_3=J_5=J_6=0$, together with the transmission matrix $T_{mn}$, we can calculate the currents $J_1 = -J_4$ and the voltages $V_2$, $V_3$, $V_5$ and $V_6$ by solving Eq.(\ref{Landauer}). Then, the longitudinal resistance $\rho_{xx} \equiv (V_2-V_3)/J_1$ and Hall resistance $\rho_{xy}\equiv (V_6-V_2)/J_1$ are obtained.

Assuming $p$ is the probability of the additional scattering shown in Fig.1(d)-(f), the transmission matrix elements of the Hall device can be written as
$T_{n,n-2}=p$ (if $n=1$ or $2$, $n-2=5$ or $6$), $T_{n,n-1}=1-p$ ($n+1=6$, if $n=1$), and others are zero.
Taking into account the boundary conditions, the Landauer-B\"{u}ttiker formula can be written as $\mathcal{J}=\mathcal{T}\mathcal{V}$ with $\mathcal{J} = (J,0,0,-J,0,0)^\dagger$, $\mathcal{V} = (1,V_2,V_3,0,V_5,V_6)^\dagger$, and
\begin{equation}
\begin{split}\mathcal{T}=\left(
\begin{array}{c c c c c c}
1 & p-1 & p & 0 & 0 & 0 \\
0 & 1 & p-1 & p & 0 & 0 \\
0 & 0 & 1 & p-1 & p & 0 \\
0 & 0 & 0 & 1 & p-1 & p \\
p & 0 & 0 & 0 & 1 & p-1 \\
p-1 & p & 0 & 0 & 0 & 1 \\
\end{array}\right)
\end{split}
\end{equation}
Solving the above linear equations $\mathcal{J}=\mathcal{T}\mathcal{V}$, we get the injecting current and the boundary voltages
\begin{equation}
\begin{split}
&V_2=\frac{p(1-p)}{1+p-p^2}, ~~V_3=\frac{p}{1+p-p^2}\\
&V_5=\frac{1}{1+p-p^2}, ~~V_6=\frac{1-p^2}{1+p-p^2}\\
&J=\frac{e^2}{h}\frac{1-p^3}{1+p-p^2}\label{V1}
\end{split}
\end{equation}
From Eq.(\ref{V1}), the longitudinal resistance and Hall resistance are expressed as
\begin{equation}
\begin{split}
&\rho_{xx}=-\frac{h}{e^2}\frac{p^2}{1-p^3}\approx -p^2\frac{h}{e^2}\\
&\rho_{xy}=\frac{h}{e^2}\frac{1-p}{1-p^3}\approx (1-p)\frac{h}{e^2}\label{rho1}
\end{split}
\end{equation}
These results suggest that, when $N_l>N_c$, $\rho_{xx}$ is almost zero for small $p$ and $\rho_{xy}$ is driven away from the quantized value. In fact, in the experiments
of QAHE in MTIs, $\rho_{xx}$ is nonzero and $\rho_{xy}$ is quantized.

(2) The thickness $N$ of terminal leads is thinner than that of the central region, i.e., $N_l<N_c$. In this case, electrons in the central region are scattered by the boundary of the Hall bar [the thick black lines in Fig.1(b)]. Similar to the case of $N_l>N_c$, additional scattering also occurs near the entrance of the Hall bar, as shown in Fig.1(b). The situation is equivalent to the case of splitting gates \cite{S.Datta.1995} shown in Fig.1(c), in which terminal 3 (6) 'sees' only the channels originating from terminal 4 (1). Hence,
\begin{equation}
V_3=V_4,~ V_6=V_1 \label{V3V6}
\end{equation}
Assuming the probability of additional scattering is $p$, i.e., $T_{26}=T_{53}=p$. Since terminal 2 (5) 'sees' channels originating from both terminal 3 (6) and terminal 6 (3) with weights of $1-p$ and $p$, respectively, we get $V_2=(1-p)V_3 + pV_6$, $V_5=(1-p)V_6+pV_3$. Combined with the conditions in Eq.(\ref{V3V6}), we have
\begin{equation}
V_2=(1-p)V_4 + pV_1,~ V_5=(1-p)V_1+pV_4 \label{V2V5}
\end{equation}
Finally, we can obtain the longitudinal bias $V_2-V_3=p(V_1-V_4)$, which means $\rho_{xx}\propto p$. In addition, the net current of terminal 1 is determined by the incoming current (to the central scattering region) and the outgoing current (from the central scattering region) in the form of $J_1=J_{1,in}-J_{1,out}$. According to Fig.1(b), $T_{26}=T_{53}=p$, $T_{23}=T_{56}=1-p$ and other $T_{mn}=1$, one easily gets
\begin{equation}
\begin{split}
&J_{1,in}=\frac{h}{e^2}\sum_n T_{n1}V_1=T_{12}V_1\\
&J_{1,out}=\frac{h}{e^2}\sum_n T_{1n}V_n=T_{12}V_2\\
&J=J_{1}=\frac{e^2}{h}(V_1-V_2)=\frac{e^2}{h}(1-p)(V_1-V_4)\label{J1}
\end{split}
\end{equation}
Then, from Eq.(\ref{V3V6}-\ref{J1}), we can derive $\rho_{xx}$ and $\rho_{xy}$ as
\begin{equation}
\rho_{xx}=\frac{h}{e^2}\frac{p}{1-p}\approx p\frac{h}{e^2}, ~~~\rho_{xy}=\frac{h}{e^2}\label{rho2}
\end{equation}
Obviously, when $N_l<N_c$, $\rho_{xy}$ is perfectly quantized and $\rho_{xx}$ is slightly dissipative. This conclusion provides a reasonable explanation to the origination of nonzero $\rho_{xx}$ and quantized $\rho_{xy}$ of QAH effect in a series of experiments performed on MTIs.

From these theoretical analysis, we have shown that additional scattering at the entrance of the Hall bar affects the QAHE in an MTI film. In the following, we will prove that, the geometric mismatch between the detecting leads and the central region induces additional scattering and consequently leads to the dissipative edge states in the QAHE of MTI films.

\section{Numerical Results and Discussions}

In this section, we present quantitative results of geometric effect on QAHE. In order to study the geometric effect of the system's thickness described by dimension $z$, in stead of the 2D TI system,\cite{PhysicalReviewLetters111.146802.Lu.2013,Phys.Rev.B89.155419.Zhang.2014} we construct a six-terminal scattering device fabricated on the magnetically doped $(Bi,Sb)_2Te_3$ film with finite thickness. In the following, we numerically calculate the longitudinal resistance $\rho_{xx}$ and the Hall resistance $\rho_{xy}$ with a 3D tight-binding Hamiltonian and non-equilibrium Green's function formalism. The $k \cdot p$ model Hamiltonian of a $(Bi,Sb)_2Te_3$ film is expressed as\cite{Phys.Rev.B82.045122.Liu.2010,NatPhys5.438-442.Zhang.2009}
\begin{equation}
H_0(k)=\epsilon_k  + M_k \sigma_0 \tau_z
+ A_\perp k_z \sigma_z \tau_x + A_\parallel (k_x \sigma_x + k_y \sigma_y) \tau_x\nonumber
\end{equation}
where, $\epsilon_k=C_0 + C_\perp k_z^2 + C_\parallel (k_x^2+k_y^2)$, and $M_k=D_0 + D_\perp k_z^2 + D_\parallel (k_x^2+k_y^2)$. $\sigma_0$ and $\tau_0$ are 2$\times$2 unitary matrices. $\sigma_{x,y,z}$ and $\tau_{x,y,z}$ are Pauli matrices, representing the real and pseudo spins formed by four low-lying states $|P1_z^+,\uparrow(\downarrow)\rangle$ and $|P2_z^-,\uparrow(\downarrow)\rangle$ at the $\Gamma$ point. Here, $\epsilon_k$ can only shift the energy band globally, and has no impact on the topological properties. So we set $\epsilon_k=0$ for an intuitive analysis. The other parameters are $A_\perp=2.2eV$\AA, $A_\parallel=4.1eV$\AA, $D_0=-0.28eV$, $D_\perp=10eV$\AA$^2$, $D_\parallel=56.6eV$\AA$^2$, respectively.\cite{NatPhys5.438-442.Zhang.2009} To observe QAH effect in a 3D TI, the magnetically doping is needed to induce the exchange field $M$ along $z$-direction. Then, the total Hamiltonian becomes $H=H_0 + M\sigma_z\tau_0$. Further, as in the experiments, an external magnetic field is applied to eliminate the residual bulk conducting states.\cite{PhysicalReviewB80.235411.Xing.2009,PhysicalReviewB89.085309.Xing.2014} Hence the vector potential induced by the magnetic field is also considered. Using the finite-difference approximation, we can get the effective tight-binding Hamiltonian of the 3D MTI system in square lattice\cite{Phys.Rev.B89.245107.Zhang.2014}
\begin{equation}
\begin{split}
H&=\sum_\mathbf{i} d_\mathbf{i}^\dagger H_\mathbf{i} d_\mathbf{i}
+ \sum_{\mathbf{i},\alpha} d_\mathbf{i}^\dagger H_\alpha d_{\mathbf{i}+\mathbf{a}_\alpha}+H.c.\label{H0}
\end{split}
\end{equation}
with
\begin{equation}
\begin{split}
H_\mathbf{i} &= \epsilon_\mathbf{i}\sigma_0\tau_0+(D_0-2\sum_\alpha\frac{D_\alpha}{a^2})\sigma_0\tau_z+M\sigma_z\tau_0\\
H_\alpha &= \frac{D_\alpha}{a^2}\sigma_0\tau_z-i\frac{A_\alpha}{2a}\sigma_\alpha\tau_x\nonumber
\end{split}
\end{equation}
where, $d_\mathbf{i}=[d_{\mathbf{i},1_z^+,\uparrow},d_{\mathbf{i},2_z^-,\uparrow},d_{\mathbf{i},1_z^+,\downarrow},d_{\mathbf{i},2_z^-,\downarrow}]$, $\alpha=x,y,z$, $F_x=F_y=F_\parallel$, $F_z=F_\perp$, $F=A,D$. $\epsilon_{\mathbf i}$ is the on-site energy induced by the random disorder.

In the numerical calculation, we set the lattice constant $a=5$\AA. By tuning the relative composition of Bi and Sb, the Fermi surface can be shifted into the bulk gap. The Fermi energy is set as $E_F=0.01eV$ that is within the bulk gap ($\approx|D_0|$). We choose other computational parameters as, the exchange field $M=0.15eV$, the widths of leads and scattering region $W_l=W_c=30a$, and the length of central scattering region $L_c=90a$. The transmission matrix elements between the six terminals are calculated through non-equilibrium Green's functions
$$T_{mn}= {\rm Tr}[\Gamma_m G^r \Gamma_n G^a ]$$
Here $'{\rm Tr}'$ denotes the trace over real space and orbital space $|P2_z^-/P1_z^+,\uparrow/\downarrow\rangle$. The linewidth function is $\Gamma_m = i[\Sigma^r_m-\Sigma^{r,\dagger}_m]$, and the retarded green's function is defined as $G^r=G^{a,\dagger}=(E-H_c-\sum_m\Sigma_m^r)^{-1}$ where $H_c$ is the Hamiltonian of the central scattering region and $\Sigma^r_m=H_{cm}g^r_m H_{m c}$ is the retarded self-energy function contributed by the $m$-th terminal lead. $g^r_m$ is the surface Green's function of the $m$th lead, which can be calculated iteratively through transfer matrix\cite{J.Phys.F:Met.Phys.1984,J.Phys.F:Met.Phys.1985} or Bloch eigenvector.\cite{Phys.Rev.B23.4988.1981,Phys.Rev.B23.4997.1981} Finally, we can get the longitudinal resistance $\rho_{xx}$ and Hall resistance $\rho_{xy}$ from the Landauer-B\"{u}ttiker formula in Eq.(\ref{Landauer}).

\begin{figure}
\includegraphics[width=8.5cm,totalheight=6cm, clip=]{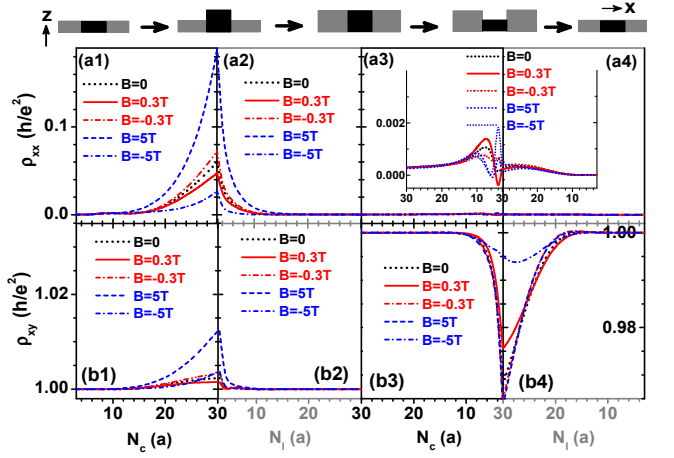}
\caption{ (Color online)
Variations of $\rho_{xx}$ and $\rho_{xy}$ with geometric structures described by lead thickness $N_l$ and central region thickness $N_c$.
The top sketches depict different geometric structures. From the left to the right, we have $N_l=N_c=5a$, $N_l<N_c$, $N_l=N_c=30a$, $N_l>N_c$ and $N_l=N_c=5a$. The inset panels are the highlights of panels (a3) and (a4).} \label{structure}
\end{figure}

In the experimental setup, the thickness of an MTI thin film is usually 5-10 quintuple layers, which is around 75-150\AA. In our numerical calculation, the thickness of leads and the central region is set as $N_l,N_c\in[15,150]$\AA. In order to study how the geometric structure in the third dimension $z$ affects the QAHE in MTI films, we explore a series of six-terminal devices with different lead thickness $N_l$ and central region thickness $N_c$, which is shown in the top sketches of Fig.2. In these sketches, from the left to the right, the geometric structure of the Hall bar evolves gradually from the slim structure ($N_l=N_c=5a$), to the convex structure ($N_l < N_c$), the thick structure ($N_l=N_c=30a$), the concave structure ($N_l>N_c$), and finally back to the slim structure ($N_l=N_c=5a$).

We first confirm the convex and concave structures are respectively corresponding to Fig.1(b) and Fig.1(d-f) from the numerical results. In the top panel, we sketch the lateral view of the six-terminal system. From the left to the right, the geometry of the system changes from thin, convex, thick to concave, and comes back to thin, corresponding to $N_l=N_c=3a$, $N_l(=3a)<N_c(=30a)$, $N_l=N_c=30a$, $N_l(=30a)>N_c(=3a)$ and $N_l=N_c=3a$, respectively. For the convex structure with $N_l=3a$ and $N_c=30a$ (the second sketch of top panels of Fig.2), the transmission coefficients are $T_{26}=T_{53}=0.04$, $T_{62}=T_{35}=0.002$. Comparing to $T_{26}$, the transmission $T_{62}$ from lead 2 to 6 can be neglected. This coincides with the scattering mechanism depicted in Fig.1(b). While for the concave structure (the fourth sketch in the top panels of the Fig.2), numerical results also confirm the scattering mechanisms shown in Fig.1(d)-(f). For instance, we find $T_{62}\gg T_{26}$, $T_{35}\gg T_{53}$, which agrees with Fig.1(d). $T_{13}$, $T_{24}$, $T_{51}$ and $T_{46}$ in the concave structure are also much greater than those in the convex structure, which agrees with Fig.1(e) and Fig.1(f).

Now, we verify that the convex structure corresponding to Fig.1(b) is responsible for the dissipative edge states of QAHE. Considering different $N_c$ and $N_l$, the six-terminal system falls into different geometry, i.e., convex geometry or concave geometry, as shown in the top panels of Fig.2. Accordingly, $\rho_{xx}$ and $\rho_{xy}$ vs $N_c$ or $N_l$ for different magnetic fields $B$ are plotted in Fig.2(a1)-(a4) and Fig.2(b1)-(b4), respectively. It is found that for the convex geometry (the left panels of Fig.2), the longitudinal resistance $\rho_{xx}$ increases promptly from zero, reaches to the maximum when $N_c$ is the largest and then decreases back to zero rapidly with the increase of $N_l$ as shown in Fig.2(a1) and Fig.2(a2). At the same time, the Hall resistance $\rho_{xy}$ increases slowly from $h/e^2$ and basically maintains quantized as shown in Fig.2(b1) and (b2). Comparing to the ideal QAHE, the deviation of the longitudinal resistance $\delta\rho_{xx}$ is an order of magnitude larger than $\delta\rho_{xy}$, which is consistent with the experimental observations.\cite{Science340.167-170.Chang.2013} Therefore, the convex structure can give rise to additional scattering which is responsible for the dissipative edge states. On the other hand, for the concave geometry, $\rho_{xx}$ is basically quantized [Fig.2(a3) and Fig.2(a4)] while $\rho_{xy}$ deviates from the quantized value significantly [Fig.2(b3) and (b4)]. The insets of panels (a3) and (a4) show that the strong scattering near the entrance of the terminals may induce a negative $\rho_{xx}$, which agrees with Eq.(\ref{rho1}). The concave geometry induces larger deviation of $\rho_{xy}$ than $\rho_{xx}$, and the leads are more prone to be destroyed during electrode building. In this sense, the convex geometry, and consequently the deviation of $\rho_{xy}$ is more common in experiments.

\begin{figure}
\includegraphics[width=8.5cm,totalheight=6.5cm, clip=]{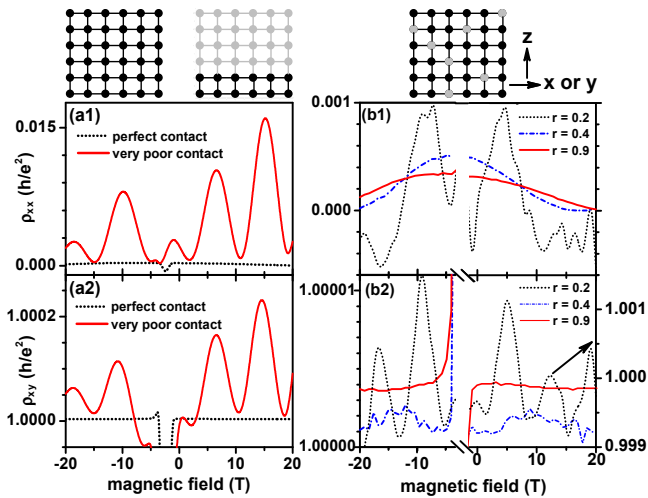}
\caption{ (Color online) $\rho_{xx}$ and $\rho_{xy}$ vary with magnetic field strength $B$ for different contact patterns: $N_c=N_l=30a$, (left sketch), $N_l=10a$, $N_c=30a$ (middle sketch) and partially connected in a random configuration with $N_c=N_l=30a$ (right sketch).}
\end{figure}

In order to get more realistic results, we consider more delicate geometric structures. In Fig.3, we plot $\rho_{xx}$ and $\rho_{xy}$ vs external magnetic field $B$ for three different contact patterns as shown in the top panels of Fig.3. When the leads and the central region are perfectly contacted, i.e., $N_c=N_l=30a$, all of the lattice sites in the lead (the black dots) are contacted to the central region (the up-left sketch). Then, the ideal QAHE is observed with $\rho_{xx}=0$ and $\rho_{xy}=h/e^2$. The perfect QAHE is kept for a broad range of magnetic fields [the black dotted lines in Fig.3(a1) and Fig.3(a2)]. We evaluate an extreme case of the mismatch between leads and the central region (the up-middle sketch in fig.3), in which the thickness of the leads (the black dots) is much smaller than that of the central region (including the grey dots), i.e., $N_l=10a$ and $N_c=30a$. Numerical results show that, additional scattering in the Hall bar induces larger nonzero $\rho_{xx}$ and the quantization of $\rho_{xy}$ is not so well [the black dotted lines in Fig.3(a1) and Fig.3(a2)]. In addition, in this poor contact situation, $\rho_{xx}$ and $\rho_{xy}$ oscillates with the variation of the magnetic field strength. The oscillation period is determined by $\Delta B\approx\frac{h/e}{W_cL_c}$. In a more general situation, the lattice sites of the lead (the black dots)through which the central region is connected, are randomly distributed, as shown in the up-right sketch in Fig.3. We define the contact ratio $r$ as that between coupled lattices and total lattices. For a very poor contact $r=0.2$, both $\rho_{xx}$ and $\rho_{xy}$ deviate from ideal values to the same extent [the black dotted lines in Fig.3(b1) and Fig.3(b2)]. With the increasing of $r$, $\rho_{xy}$ approaches to the quantized value $h/e^2$ gradually, while $\rho_{xx}$ is roughly kept in the same order as in the case of $r=0.2$. This fact shows that the geometric structure has a remarkable impact on $\rho_{xx}$, but almost does not affect $\rho_{xy}$.

Next, we focus on the influence of the magnetic field. As demonstrated in experiments\cite{Science340.167-170.Chang.2013,PhysicalReviewLetters113.Kou.2014,PhysicalReviewLetters114.187201.Bestwick.2015,NatureMaterials14.473-477.Chang.2015},
the Hall resistance $\rho_{xy}$ is quantized while the longitudinal resistance $\rho_{xx}$ is nonzero in zero or weak magnetic fields, i.e., the edge states are slightly dissipative. The only way to eliminate this dissipation is applying an external magnetic field. A strong magnetic field ($|B|>5T$) finally realizes the ideal QAHE.\cite{Science340.167-170.Chang.2013} However, in some cases, $\rho_{xx}$ does not go to zero even in a strong magnetic field of $10T$.\cite{PhysicalReviewLetters113.Kou.2014} Here, we attribute the nonzero $\rho_{xx}$ to the non-ideal contact between metallic leads and the central sample.

From Fig.3(b), we can find that $\rho_{xx}$ is nonzero at $B=0$. At relatively good contact ($r=0.4$ and $0.9$), with the increasing of $B$, $\rho_{xx}$ is quickly depressed to zero, while $\rho_{xy}$ just slightly oscillates. Our results confirm the experimental observation that $\rho_{xx}$ is much more sensitive to the magnetic field than $\rho_{xy}$. For a moderate contact with $r=0.4$ [blue curve in Fig.3(b1)], $\rho_{xx}$ quickly drops to zero as the magnetic field is increased, which is in good agreement with the observation of Ref.\citenum{Science340.167-170.Chang.2013}. In contrast, for a good contact [$r=0.9$, red curve in Fig.3(b1)] the dissipative component $\rho_{xx}$ is smaller but it goes to zero at a slower rate in the presence of magnetic field, as in Ref.\citenum{PhysicalReviewLetters113.Kou.2014}. In addition, our numerical data show that $\rho_{xy}$ has small oscillations as a function of the magnetic field which was also observed in experiments. It should be noted that, the chirality of the edge states in the QAH system and the quantum Hall system is opposite in the presence of a negative magnetic field. Therefore, when the magnetic field is reversed from positive to negative ($-3T<B<0$ region in Fig.3(b)), $\rho_{xx}$ and $\rho_{xy}$ are singular. The same phenomena have also been observed in experiments.\cite{Science340.167-170.Chang.2013,PhysicalReviewLetters113.Kou.2014}

\begin{figure}[t]
\includegraphics[width=6cm, clip=]{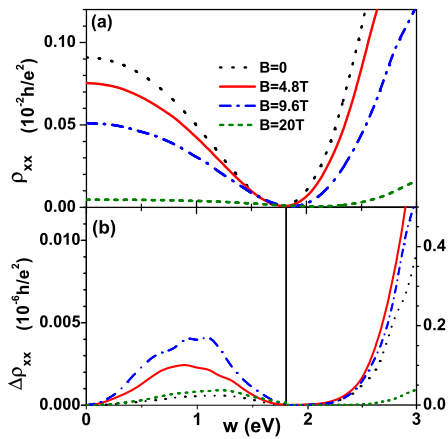}
\caption{ (Color online) $\rho_{xx}$ and its fluctuation $\Delta\rho_{xx}$ vs the disorder strength $w$ at $N_l=N_c=10a$ for different magnetic fields.}
\end{figure}

Finally, the disorder effect is evaluated, since magnetic doping on MTIs can induce static disorder. The Anderson-type static disorder is modeled as the random on-site energy, which uniformly distributes in the interval $[-w/2,w/2]$ with $w$ the disorder strength. Here, we consider the perfect contact situation, i.e., $N_c=N_l=10a$. Numerical results show that in this case the Hall resistance $\rho_{xy}$ is hardly affected by the Anderson-type disorder. Therefore, in Fig.4, we plot only the longitudinal resistance $\rho_{xx}$ and its fluctuation $\Delta\rho_{xx}$ vs the disorder strength $w$ for different magnetic fields $B$. It is found that, moderate disorder can induce the localization of residual bulk states and suppress the longitudinal resistance $\rho_{xx}$, while the edge states remain intact. Meanwhile, the fluctuation of $\rho_{xx}$ is slightly enhanced [left part of Fig.4(b)]. Near $w=1.8eV$, an ideal QAHE occurs with zero fluctuation of $\rho_{xx}$. This is very similar to topological Anderson insulators in quantum spin Hall effect.\cite{PhysicalReviewLetters102.Li.2009,PhysicalReviewB84.Xing.2011} When the disorder is further increased, the backscattering between the chiral edge states located on the opposite edges will destroy the QAH states. As a result, $\rho_{xx}$ and $\Delta\rho_{xx}$ abruptly increases. Our numerical results show that, the chiral edge states are immune to weak disorder. Moreover, appropriate disorder is beneficial for the QAHE by suppressing the dissipative edge states, which provides useful insight on future experiments.

\section{Summary} 
In summary, we expound the dissipative edge states of QAHE in MTIs with simple and intuitive physics. We have shown by both theoretical analysis and numerical evidence that, due to the geometric mismatch between the leads and the central sample, additional scattering is induced in the Hall bar and gives rise to the nonzero longitudinal resistance $\rho_{xx}$ and quantized Hall resistance $\rho_{xy}$. Either the convex geometric structure or poor contact between leads and the central region can contribute to the geometric mismatch. It is numerically found that, the nonzero $\rho_{xx}$ can be effectively suppressed by external magnetic fields, which agrees with experimental observations. Besides, Anderson-type disorder is also shown to be beneficial in suppressing the dissipation of $\rho_{xx}$. These findings provide useful insight to the design and application of future QAHE devices.

$${\bf ACKNOWLEDGMENTS}$$
This work is supported by the MOST Project of China (2017YFA0303301, No. 2016YFA0300603, No. 2015CB921102 and No. 2014CB920903), and the NNSFC (No.11674024, No.11504240, No.11574029, No.11574007).


\end{document}